\documentstyle[preprint,aps,epsf,floats]{revtex}

\newcommand{\beq}{\begin{equation}}
\newcommand{\eeq}{\end{equation}}
\newcommand{\bea}{\begin{eqlanarray}}
\newcommand{\eea}{\end{eqnarray}}
\newcommand{\ben}{\begin{eqnarray*}}
\newcommand{\een}{\end{eqnarray*}}

\newcommand{\simlt}{\stackrel{<}{{}_\sim}}

\tightenlines

\begin{document}

\title{Range Corrections to Doublet S-Wave Neutron-Deuteron Scattering}

\author{H.-W. Hammer\footnote{hammer@mps.ohio-state.edu} 
and Thomas Mehen\footnote{mehen@mps.ohio-state.edu}}

\address{Department of Physics\\ The Ohio State University, Columbus, OH 
43210}

\date{May 2001}  

\maketitle

\begin{abstract}
We calculate the range corrections to S-wave neutron-deuteron scattering
in the doublet channel ($S=1/2$) to first order in $r/a$ where
$a$ is the scattering length and $r$ the effective range.
Ultraviolet divergences appearing at this order 
can be absorbed into a redefinition of the leading order three-body force. 
The corrections to the elastic scattering amplitude below 
the deuteron breakup threshold are computed. Inclusion of the 
range corrections gives good agreement with measured
scattering data and potential model calculations. 
\end{abstract}

\bigskip
\pacs{PACS number(s): }

\newpage                  

There has been much interest recently in applying Effective Field Theory (EFT)
methods to nuclear physics \cite{EFT98,EFT99,Birareview,BEANE99}. EFT provides
a framework in which to exploit the separation of scales in physical systems in
order to perform systematic, model-independent calculations. For nuclear
few-body systems, the long-distance scale is 
set by  the large two-body scattering lengths, while the short-distance 
scale is set by the range of the nuclear force. 
The EFT includes long-distance physics explicitly, while
corrections from short-distance physics are calculated perturbatively in an
expansion in the ratio of these two scales. In the two-body system for
small momenta ($p \simlt m_\pi$), this program has been very successful and 
calculations of deuteron properties and electroweak
processes have been carried out to $1\%$ accuracy (for a recent review,
see Ref. \cite{BEANE99}).

In the nuclear three-body system, considerable progress has been made as 
well. In most three-body channels, the two-body EFT can  be extended 
in a straightforward way \cite{BvK98,BHK98,BeG00,GBG00}. 
However, the  S-wave in the doublet channel
($S=1/2$) of neutron-deuteron scattering (with the triton as a three-body bound
state) is more complicated and exhibits some surprising phenomena
\cite{Tho35,Efi71,Phi68}. 
The renormalization of the three-body equations requires
a one-parameter three-body force at leading order whose renormalization group
evolution is governed by a limit cycle \cite{BHK00}. The variation of the
three-body force parameter gives a compelling explanation of the 
Phillips line (an essentially equivalent explanation was previously 
given in Refs. \cite{Dan61,EfT85}). 
The phase shifts for S-wave neutron-deuteron scattering in
the doublet channel have been studied at order $(r/a)^0$, where $a$ is the
scattering length and $r$ the effective range \cite{BHK00}. In this paper, we
calculate the linear corrections in $r/a$ to the elastic 
scattering phase shifts below the deuteron breakup threshold. 
The linear range corrections to the Phillips
line have been studied in Ref. \cite{EfT85} using a different formalism.

Elastic neutron-deuteron scattering below the threshold of deuteron breakup
can be described by an effective Lagrangian that includes only nucleons 
and has no explicit pions. 
For three-body calculations, it is convenient to use the dibaryon
formalism of Ref. \cite{Kap97} in which auxiliary fields with baryon number 
two are introduced to represent two-nucleon states in a given partial wave. 
The effective Lagrangian for the nucleon-deuteron system is \cite{BHK00} 
\begin{eqnarray}
\label{Lag}
{\cal L} &=& 
N^{\dagger}\bigg( i \partial_0 + {\vec{\nabla}^2 \over 2 M} \bigg) N 
- t_l^{\dagger} \bigg( i \partial_0 + \frac{\vec{\nabla}^2}{4 M} 
   - \Delta_t \bigg) t_l 
- s_m^{\dagger} \bigg( i \partial_0 + {\vec{\nabla}^2 \over 4 M} 
  -\Delta_s \bigg) s_m \\
&&- {g_t\over 2} \Big(t_l^{\dagger} N^T \tau_2 \sigma_l 
    \sigma_2 N + \mbox{h.c.}\Big)
- {g_s \over 2} \Big(s_m^\dagger N^T \sigma_2 \tau_m \tau_2 N 
   + \mbox{h.c.} \Big)  \, \nonumber \\
&& -G_3\,N^\dagger \bigg( g_t^2
( t_l \sigma_l )^\dagger t_{l'} \sigma_{l'}
+\frac{1}{3}\,g_t g_s \left[ (t_l \sigma_l)^\dagger 
  s_m \tau_m + \mbox{h.c.} \right] 
+ g_s^2 (s_m \tau_m )^\dagger s_{m'} \tau_{m'} \bigg) N +\ldots\nonumber\,,
\end{eqnarray}
where $N$ represents the nucleon field and $t_l$ ($s_m$) are 
the dibaryon fields for the $^3S_1$ ($^1S_0$) channels and
carry spin (isospin) one, respectively. The dots indicate higher order
terms with more fields/derivatives, which do not contribute to the
order we are working.
The first line in Eq.~(\ref{Lag}) contains the kinetic terms for the fields
$N$, $t_l$, and $s_m$.
The second line gives the coupling of the dibaryon fields 
to nucleon fields where $\sigma_l$ $(\tau_m)$
are Pauli matrices acting in spin (isospin) space. 
Finally, the third line contains the three-body force. 
Note that this is the only three-body operator without derivatives
that preserves spin and isospin symmetry \cite{BHK00,MSW99}. Other
non-derivative three-body operators can be related to the one shown via
Fierz transformations.
Since the action is quadratic in the fields $t_l$ and $s_m$, it is
straightforward to integrate them out and show that the theory is equivalent
to one of nonrelativistic nucleons interacting via two-body and three-body
contact interactions \cite{BeG00,GBG00,BHK00}.

\begin{figure}[t]
  \vspace*{0.2cm}
  \centerline{\epsfxsize=12.0cm \epsfbox{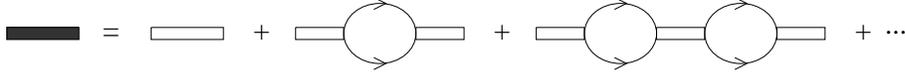}  }
  \vspace*{0.2cm}
\caption{Geometric series leading to the exact dibaryon propagator.
Double (single) lines represent bare dibaryon (nucleon) propagators,
respectively.}
\label{resum}
\end{figure}

To obtain the exact dibaryon propagator, the bare propagator must be
dressed by nucleon loops to all orders. This is illustrated in
Fig.~\ref{resum}. 
The diagrams form a geometric series and are easily summed; the result is
\begin{eqnarray}
\label{prop} 
i D_{j}(q_0,\vec{q}\,) &=&  
  {-i \; 2 \pi/( M g_{j}^2) \over -\gamma_{j} + \frac{1}{2} r_j 
     \left( \gamma_j^2 +M q_0 - 
    \vec{q}^{\,2}/4 \right)+ \sqrt{-M q_0 + \vec{q}^{\,2}/4 -i \epsilon} }
\,,
\end{eqnarray} 
where the subscript $j=t\,(s)$ for the $^3S_1\,(^1S_0)$ channel.
This propagator has a pole  at $M q_0 - \vec{q}^{\,2}/4 =
- \gamma_j^2$, where  $\gamma_t=45.68$ MeV and $\gamma_s=-7.88$ MeV.  The 
effective ranges are $r_t=1.76$ fm and $r_s=2.75$ fm. These effective range 
parameters are related to the parameters appearing in  the 
Lagrangian via
\begin{eqnarray} 
{\gamma_j} = {M^2 g_{j}^2 \over 4\pi}\left(1-\sqrt{1-{16\pi^2 \Delta_j
\over M^3 g_j^4}}\;\right)\qquad\mbox{ and }\qquad
r_{j}= {4 \pi \over M^2 g_{j}^2} \, . 
\end{eqnarray} 
The scattering length $a_j$ is given by
$-1/a_j =-\gamma_j +r_j \gamma_j^2/2$.
Thus the Lagrangian of Eq.~(\ref{Lag}) reproduces the first two
terms in the effective range expansion of the two-body
scattering amplitude. Note that including 
the kinetic terms for the dibaryon fields in Eq.~(\ref{Lag})
is necessary to obtain the range correction. 

The power counting in an EFT makes it possible
to organize calculations in a
systematic expansion in a small parameter. In the two-body sector, the 
power-counting  scheme of \cite{KSW98,UvK99} takes $p \sim 1/a \sim Q$,  
where $p$ is the typical momentum of a nucleon and $1/a$ is the 
inverse scattering length. Note that $\gamma$ is $O(Q)$ as well since 
$\gamma = 1/a+O(r/a^2)$.  The expansion parameter
of the theory is $Q/\Lambda$, where $\Lambda$ represents the scale where
short-distance physics becomes important. 
For systems interacting via short-range
interactions, the effective range is expected to be set by $\Lambda$. So the
expansion parameter is $Q/\Lambda \sim  r/a \approx\gamma r$.\footnote{Note,
that it is sufficient to take $\gamma_t r_t =\gamma_s r_s =\gamma r$ 
for power counting purposes.} The leading order term
in the expansion of the two-body scattering amplitude is  $O(Q^{-1})$.
If we integrate out the dibaryon fields, the two-body operators without 
derivatives are treated nonperturbatively because the renormalization 
group equations  dictate that their coefficients are $O(Q^{-1})$.
In the dibaryon formalism, $2\pi\Delta/(Mg^2 )=1/a \sim Q$, which
requires summing the bubbles in Fig.~\ref{resum}.  
At leading order in $Q$, the theory reproduces Eq.~(\ref{prop}) in the 
limit $r_t,r_{s} \rightarrow 0$.  The two-body operators with two spatial 
derivatives or one time derivative which are given by the kinetic terms 
of the dibaryon fields  appear at next-to-leading order in the $Q$ expansion, 
giving a contribution of $O(Q^0)$ to the scattering amplitude. When
these operators are included in the three-body problem, they also give a
correction that is  suppressed by one power of $Q$ relative to the leading
order.   Since the coefficients of these operators are proportional to the
two-body  effective ranges, we refer to these corrections as effective range 
corrections.

\begin{figure}[t]
  \vspace*{0.5cm}
  \centerline{\epsfxsize=14.0cm \epsfbox{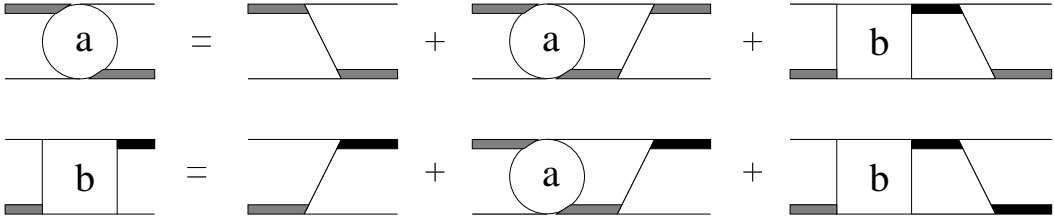}  }
  \vspace*{0.3cm}
\caption{Leading order integral equation. Shaded (full) double
lines indicate spin-triplet (spin-singlet) dibaryon. Single line
indicates nucleon propagator.}
\label{int}
\end{figure}

Next we consider power counting in the three-body sector. EFT power counting
shows that all diagrams that contain only non-derivative contact 
interactions are $O(Q^{-2})$ \cite{BHK00}.  These diagrams can be 
summed using the integral equation shown in  Fig.~\ref{int}. 
At this order in the EFT, it is appropriate 
to use the propagator of Eq.~(\ref{prop}) in the limit $r_t,r_s \to 0$. 
However, it turns out that the resulting integral equation has no unique
solution if the cutoff $\Lambda$ is taken to infinity \cite{Dan61,STM57}. 
If the integral equation is regularized with a finite cutoff $\Lambda$, 
the solution displays a
strong cutoff dependence. The integral equation can be renormalized
by absorbing the cutoff dependence into the three-body force
of Eq.~(\ref{Lag}) \cite{BHK00}. 
The diagrams with the three-body force are naively 
$O(Q^0)$.  However, in the $S=1/2$ channel, the leading three-body
operator with no derivatives is relevant at low energies
because the renormalization group evolution enhances the coefficient $G_3$ 
in Eq.~(\ref{Lag}) to $O(Q^{-2})$ rather than $O(Q^0)$ as naively expected 
\cite{BHK00}.\footnote{Note that the three-body force in Eq.~(\ref{Lag}) 
does not contribute in the quartet channel because of the  Pauli 
exclusion principle.}  
It is convenient to pull out a factor of $2M/\Lambda^2$ and
define $G_3 (\Lambda)=2M H (\Lambda)/\Lambda^2$ with
\beq
\label{runH}
H (\Lambda)=-    \frac{\sin[s_0\ln({\Lambda}/{\Lambda_*})-
                   {\rm arctg}(1/s_0)]}
                 {\sin[s_0 \ln({\Lambda}/{\Lambda_*})+
                   {\rm arctg}(1/s_0)]}\,,
%\approx -\tan(s_0\ln({\Lambda}/{\Lambda_*})-\pi/4)\, , 
\eeq            
where
$s_0\approx 1.0064$ is determined by the asymptotic behavior of the integral
equation and $\Lambda_*$ is the three-body force parameter \cite{BHK00,BHK99}.
As a consequence, there are certain cutoffs for which the three-body
force vanishes. Since all observables  are independent of the cutoff,
it is possible to obtain a renormalized equation by choosing a cutoff
with vanishing three-body force \cite{HaM00}. The parameter $\Lambda_*$
then appears in the upper limit of the integral. Evaluating the 
diagrams in Fig.~\ref{int}, the renormalized equation in the 
limit $r_t, r_s \to 0$ takes the form \cite{BHK00,STM57}
\begin{eqnarray}
\label{inteq}
2 {-\gamma_t +\sqrt{3 p^2/4 -M E_k} \over p^2 -k^2} a^0_k(p) 
= K_k(p,k) + {2\over \pi} \int_0^{\Lambda_n(\Lambda_*)} \!\!\!\!\!\!\!
dq {q^2\,{\cal P}
\over q^2-k^2} K_k(p,q) \bigg(a^0_k(q) +3 b^0_k(q) \bigg) \\
2 {-\gamma_s +\sqrt{3 p^2/4 -M E_k} \over p^2 -k^2} b^0_k(p) 
= 3 K_k(p,k) + {2\over \pi} \int_0^{\Lambda_n(\Lambda_*)}  \!\!\!\!\!\!\!
dq {q^2\, {\cal P} 
\over q^2-k^2} K_k(p,q)\bigg(3 a^0_k(q) + b^0_k(q) \bigg)\nonumber\,, 
\end{eqnarray}
where $k$ $(p)$ denote the incoming (outgoing) momenta in the 
center-of-mass  frame, $ME_k = 3k^2/4-\gamma_t^2$ is the total energy, 
and $\Lambda_n=\Lambda_*\exp[(n\pi+\arctan(1/s_0))/s_0]$ with $n$ a 
natural number. Three-body observables are independent of $n$ up to
corrections that are suppressed by inverse powers of $\Lambda_n$ \cite{HaM00}. 
The kernel $K_k(p,q)$ arises from the S-wave projected one-nucleon  
exchange and is given by
\begin{eqnarray}
K_k(p,q) = {1\over 2 p q}\ln \left({q^2 + p q + p^2 - M E_k \over 
q^2 - p q + p^2 - M E_k}\right) \,.
\end{eqnarray}
The amplitude $a^0_k(p)$ is normalized such that $a^0_k(k) =1/(k\cot\delta)$
with $\delta$ the elastic scattering phase shift.

Recently, various authors have suggested treating range corrections
nonperturbatively in both the two- and three-body systems
\cite{Co99,CoH99,KS99,BS00,Gab01}. This can be  motivated by arguing that a
nonperturbative treatment of range corrections resums large corrections
proportional to $(\gamma_t r_t)^n$ to all orders in $n$. Since $\gamma_t r_t
\approx 0.4$, these corrections can be numerically important despite being
formally subleading in the $Q$ expansion\cite{BS00}. Alternatively, one can
imagine a power counting in which $r$ is $O(Q^{-1})$
\cite{Co99,CoH99,KS99,BS00}. In the two-body sector, such a power counting was
shown to simplify calculations and improve the convergence of the expansion
\cite{BS00}. Furthermore, the coefficients of higher derivative S-wave
operators are no longer enhanced by renormalization group evolution and naive
dimensional analysis can be used to estimate their contribution to amplitudes.

A nonperturbative treatment of range corrections in the three-body problem was
studied in Refs.~\cite{Gab01,Sim88}. The integral equation of Fig.~\ref{int} 
is solved using the propagator of Eq.~(\ref{prop}) without expanding in $r$.
This drastically changes the nature of the solution to the integral equation.
Since the dibaryon propagator falls as $1/q^{2}$ rather than $1/q$ for
large  $q$, the kernel is damped at large loop momenta, and the
integral equation has a unique solution even in the absence of the three-body
force. The three-body force is not enhanced by renormalization and is a
subleading effect suppressed by $Q^2$. There is no three-body parameter
in the leading order calculation and therefore no Phillips line.
Surprisingly, the Phillips line is not even recovered at $O(Q^0)$, 
when the three-body force is included. In Ref. \cite{Gab01}, it
was shown that when the range is treated nonperturbatively and the cutoff,
$\Lambda$, is taken to infinity, the solution is completely insensitive to the
numerical value of the three-body force. Furthermore, the obtained
scattering phase shifts strongly disagree with experiment \cite{Gab01}.

It is not clear why the nonperturbative treatment of the range corrections 
fails. One possible problem is that the dibaryon propagators have spurious
poles at $M q_0 - \vec{q}^{\,2}/4 = 2/r+O(\gamma r)$.
These poles can be avoided by using a momentum cutoff in the range 
$\gamma < \Lambda \leq  1/r$. The three-body force can be tuned in such a 
way as to leave results approximately cutoff independent when the cutoff 
is varied within this window. Unfortunately, since $a_t \approx 3 \, r_t$, 
unreasonably small cutoffs must be used in this method. 
Moreover, the nonperturbative range correction worsens the agreement with 
the Phillips line.

In the present paper, we take a different approach
and compute the range corrections to $a^0_k(k)$ perturbatively,
working to $O(Q^{-1})$. An important question is whether higher
derivative three-body operators also contribute at this order. These
operators will include at least one time derivative or two spatial
derivatives and hence contribute to the amplitude at $O(Q^2)$, according
to naive dimensional analysis.  This is much higher order than the range
correction to the three-body amplitude, which is $O(Q^{-1})$. Higher
derivative three-body operators could only contribute at this order if
there were a renormalization group enhancement of their coefficients,
which is not the case. Consequently, the  range correction is the only
contribution at this order. In the following, it is more convenient
to label the contributions by powers of $\gamma r$ relative to the 
leading order. The linear range correction is then
$O(\gamma r)$. Below, it is demonstrated that the range
correction can be calculated without introducing any terms not already
present in Eq.~(\ref{Lag}). The operator which renormalizes the leading
order calculation can also be used to renormalize the loop graphs that
appear at $O(\gamma r)$. However, the running of the coefficient $H(\Lambda)$
needs to be modified. 

Since the computation of the range corrections requires 
no additional counterterms, only the two-body 
effective ranges $r_t$ and $r_s$ enter as  new parameters at
this order of the calculation.  The only parameter to be fixed from a
three-body datum is the three-body  force parameter, $\Lambda_*$.  For
example,  if $\Lambda_*$ is fixed to reproduce
the observed neutron-deuteron scattering length, the energy dependence of
the scattering amplitude is completely predicted up to corrections of
$O((\gamma r)^2)$. 

To calculate the range corrections it is convenient to formally expand
the amplitudes $a_k(p)$, $b_k(p)$ and the three-body force $H(\Lambda)$
in powers of $\gamma r$,
\begin{eqnarray}
\label{gamrexp}
a(q) = a_k^0(q) + a_k^1(q) + ... \,,\\
b(q) = b_k^0(q) + b_k^1(q) + ... \, ,\nonumber\\ 
H(\Lambda) = H^0(\Lambda) + H^1(\Lambda)+... \nonumber\, , 
\end{eqnarray}
where the superscript denotes the power of $\gamma r$.
The amplitudes $a_k^0(q)$ and $b_k^0(q)$ are the
solutions of Eq.~(\ref{inteq}) and $H^0 (\Lambda)$ is given in
Eq.~(\ref{runH}). In the following,
we will calculate $a_k^1(k)$ and  obtain the 
renormalization group evolution of $H^1(\Lambda)$.

\begin{figure}[t]
  \vspace*{0.3cm}
  \centerline{\epsfxsize=12.0cm \epsfbox{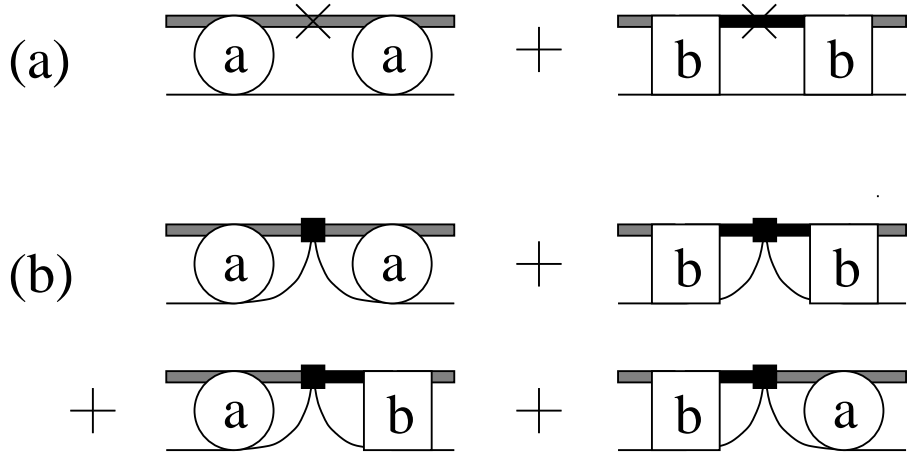}  }
  \vspace*{0.3cm}
\caption{Feynman diagrams for range correction. (a) Range correction: double
line with cross denotes $O(\gamma r)$ piece of the dibaryon propagator. 
(b) Subleading three-body force: filled square denotes insertion of 
$H^1(\Lambda)$.  The propagators are as in Fig.~\ref{int}. Not shown are 
diagrams that  vanish as $\Lambda \to \infty$.}
\label{rdiag}
\end{figure}

The Feynman diagrams contributing to $a^1_k(k)$ are shown in Fig.~\ref{rdiag}.
Fig.~\ref{rdiag}(a) shows the diagrams from the range correction.
The double lines with the cross denote the $O(\gamma r)$ 
piece of the dibaryon propagator, $\widetilde{D}_j (q_0,\vec{q}\,)$. 
Expanding Eq.~(\ref{prop}) in $r_j$, we obtain the Feynman rule
\begin{eqnarray}
\label{trick}
i\widetilde{D}_{j}(q_0,\vec{q}\,) &=& 
{-2\pi i \over M g_{j}^2} \;{r_{j} \over 2} { \left( \vec{q}^{\,2}/4
 -M q_0 - \gamma_j^2 \right) \over \left( -\gamma_{j} +
\sqrt{-M q_0 + \vec{q}^{\,2}/4 - i\epsilon}\right)^2 }\,\\
&=&{-2\pi i \over M g_{j}^2} 
\left({\gamma_j r_j \over -\gamma_{j} +
\sqrt{-M q_0 + \vec{q}^{\,2}/4 - i\epsilon} } +  {r_{j} \over 2} \right)
\,.\nonumber
\end{eqnarray}
In order to evaluate the Feynman diagrams in Fig.~\ref{rdiag}(a), it
is convenient to use the form given in the second line of Eq.~(\ref{trick}).
The constant term in Eq.~(\ref{trick}) does not contribute because the contour
integral over the zero component of the loop momentum can be evaluated without
enclosing a singularity \cite{BHK99}.
Fig.~\ref{rdiag}(b) gives the contribution of the subleading 
piece of the three-body force, $H^1(\Lambda)$.
In addition to the diagrams shown in Fig.~\ref{rdiag}(b), there are four more
diagrams where the $H^1(\Lambda)$ insertion is dressed on only one side by
either $a^0_k(q)$ or $b^0_k(q)$ and a diagram with the undressed
$H^1(\Lambda)$. These diagrams are also included in the calculation, but 
it is shown below that their contribution vanishes as $\Lambda \to \infty$.

The evaluation of these diagrams is straightforward. 
After projecting onto the S-waves, we obtain:
\begin{eqnarray}
\label{ans}
a^1_k(k) &=& \gamma_t r_t \, a^0_k(k)+ \gamma_t {2 \over \pi} 
\int_0^{\Lambda_n(\Lambda_*)} dq \frac{q^2\,{\cal P}} {q^2-k^2}\,
\frac{2 \gamma_t r_t\;a^0_k(q)^2}{\gamma_t+\sqrt{3q^2/4-ME_k}} \\
& &+ \gamma_t {2 \over \pi} \int_0^{\Lambda_n(\Lambda_*)} dq
\frac{q^2\;b^0_k(q)^2}{(q^2-k^2)^2} \;\frac{8}{3} \gamma_s r_s 
\left(-\gamma_s +\sqrt{3q^2/4-ME_k} \right)\nonumber \\
&& + {8 \gamma_t \over 3} {H^1(\Lambda_n) \over \Lambda_n^2} 
\left[ 1+{2\over \pi} \int_0^{\Lambda_n(\Lambda_*)}  
dq {q^2 \,{\cal P}\over q^2 -k^2} \bigg(a^0_k(q)+b^0_k(q) \bigg)\right]^2  
\, .\nonumber 
\end{eqnarray}
Note that for the integral in the second line no principal value prescription
is required since $b^0_k(k)=0$.

There are four contributions to the $O(\gamma r)$ corrections to the 
amplitude. First, the residue of the pole in the deuteron propagator 
changes by a factor $1+\gamma_t r_t$. This changes the LSZ factor that 
goes into the  leading-order calculation, giving rise to the first term 
in Eq.~(\ref{ans}).  The second and third terms in Eq.~(\ref{ans}) come  
from the diagrams in Fig.~\ref{rdiag}(a). The fourth term
comes from diagrams involving the subleading piece of the three-body force,
$H^1(\Lambda)$. 

In order to determine $H^1(\Lambda)$, it is necessary to know
the large $\Lambda$ dependence of the integrals in Eq.~(\ref{ans}). 
Since the combination $a^0_k(q)-b^0_k(q)$ falls off sufficiently
fast for large $q$ (cf. Ref. \cite{BHK00}), 
only the asymptotic form of the combination $a^0_k(q)+b^0_k(q)$ 
for $q \gg \gamma$ is needed:
\begin{eqnarray}
\label{asymp}
a^0_k(q)+b^0_k(q) \rightarrow s_k(q) = {\cal N}(k) 
\cos(s_0 \ln(q/\Lambda_*) +\delta) \,,
\end{eqnarray} 
where we have suppressed the dependence of ${\cal N}(k)$ on $\Lambda$
and $\gamma$.\footnote{This is can be done safely because ${\cal N}(k)$
is of $O(\Lambda^0)$ \protect\cite{BHK99}.}
It is especially important to note that the phase of $s_k(q)$ is
independent of $k$ so $s_k(q)$ factorizes into a product of  an 
unknown function of
$k$ and a known function of $q$.  This will allow us to determine
$H^1(\Lambda)$ analytically and show that all divergences in the range
correction are cancelled by this counterterm.  

Inserting Eq.~(\ref{asymp})
into Eq.~(\ref{ans}) gives the divergent piece of the range  correction
\begin{eqnarray}
\delta  a_{\rm div}  = {2 \gamma_t \over \pi}\left({\gamma_t r_t+\gamma_s r_s
 \over \sqrt{3}} \int^\Lambda {dq \over q} s^2_k(q) +{16 \over 3 \pi}
{H^1(\Lambda)\over\Lambda^2} \left[ \int^\Lambda dq s_k(q) \right]^2
\right)\,.
\end{eqnarray}
The divergence is cancelled if
\begin{eqnarray}
\label{H1}
{H^1(\Lambda)\over\Lambda^2} &=& - {\sqrt{3} \pi \over 16}
(\gamma_t r_t+\gamma_s r_s)
{\int^\Lambda dq s^2_k(q)/q \over [\int^\Lambda dq s_k(q)]^2 } \\
&=& - {\sqrt{3} \pi \over 16} (\gamma_t r_t+\gamma_s r_s)
{(1+s_0^2)^2 \over 4 s_0 \Lambda^2}
{2 s_0 \ln(\Lambda/\Lambda_*) + 2\delta + \sin(2 s_0
\ln(\Lambda/\Lambda_*)
+ 2 \delta) \over 
[\cos(s_0 \ln(\Lambda/\Lambda_*)+\delta) + s_0
\sin(s_0\ln(\Lambda/\Lambda_*)+\delta)]^2}
\nonumber
\\
&\equiv& - {\sqrt{3} \pi \over 16} (\gamma_t r_t+\gamma_s r_s){F(\Lambda) 
\over \Lambda^2} \, , \nonumber
\end{eqnarray}
where the last line defines $F(\Lambda)$. Note that $H^1(\Lambda)$ is the
coefficient of an operator with no  derivatives and cannot be a function of
$k$. As required, all $k$ dependence has
cancelled in the expression for $H^1(\Lambda)$. 
Furthermore, $H^1(\Lambda)/\Lambda^2$ scales like  $1/\Lambda^2$ (up to
logarithmic corrections) for large $\Lambda$. In the third line 
of Eq.~(\ref{ans}), it is only necessary to keep the term
in which $H^1(\Lambda)$ is multiplied by two linearly
divergent integrals. This term corresponds to the diagrams
shown in Fig.~\ref{rdiag}(b). The other terms can be discarded since
their contribution can be made arbitrarily small by choosing an 
appropriate value for $\Lambda$.
Thus the renormalized expression for the range correction is 
\begin{eqnarray}
\label{renans}
a^1_k(k) &=& \gamma_t r_t \, a^0_k(k)+ \gamma_t {2 \over \pi} 
\int_0^{\Lambda_n(\Lambda_*)} dq \frac{q^2\,{\cal P}} {q^2-k^2}\,
\frac{2 \gamma_t r_t\;a^0_k(q)^2}{\gamma_t+\sqrt{3q^2/4-ME_k}} \\
& &+ \gamma_t {2 \over \pi} \int_0^{\Lambda_n(\Lambda_*)}dq 
\frac{q^2\;b^0_k(q)^2}{(q^2-k^2)^2} \;\frac{8}{3} \gamma_s r_s \left(-\gamma_s
+\sqrt{3q^2/4-ME_k}\right)\nonumber \\
&& - {2 \gamma_t \over \sqrt{3} \pi}(\gamma_t r_t+ \gamma_s r_s) 
{F(\Lambda_n) \over \Lambda_n^2} \left[\int_0^{\Lambda_n(\Lambda_*)}  
dq \bigg(a^0_k(q)+b^0_k(q) \bigg)\right]^2  
\, .\nonumber 
\end{eqnarray}
The above expression is cutoff independent (up to corrections of
$O(1/\Lambda_n)$). Note the function $F(\Lambda)$ is known
analytically but depends on the asymptotic phase $\delta$, 
which must be determined numerically by fitting to the leading-order
solution $a^0_k(q)+b^0_k (q)$.

\begin{figure}[t]
  \vspace*{0.3cm}
  \centerline{\epsfxsize=13.0cm \epsfbox{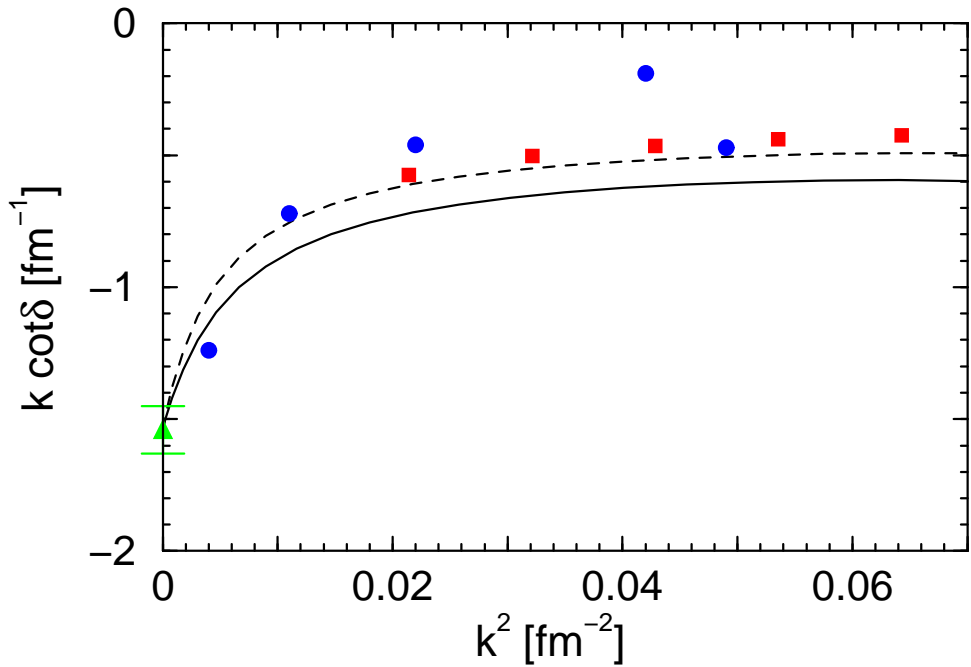}  }
  \vspace*{0.3cm}
\caption{Results for $k\cot\delta$: LO (NLO) calculation is indicated by
solid (dashed) line. Triangle gives the experimental value of
the scattering length from Ref.~\protect\cite{DKN71},
circles show the phase shift analysis of Ref.~\protect\cite{OS67}, 
and squares show the potential model calculation of 
Ref.~\protect\cite{KRT96}.}
\label{kcotnlo}
\end{figure}

In Fig.~\ref{kcotnlo}, we show our result for
$k\cot\delta$. The leading-order (LO) calculation is indicated by the 
full line, while the next-to-leading-order (NLO) calculation is given 
by the dashed line. At each order, $\Lambda_*$ is tuned to produce the 
measured neutron-deuteron scattering length,
$a_{nd}^{(1/2)}=(0.65\pm 0.04)$ fm \cite{DKN71}. 
We find $\Lambda_*^{LO} = 3.6\,\gamma_t = 0.83 \, \rm{fm}^{-1}$ 
and $\Lambda_*^{NLO} = 4.1\, \gamma_t = 0.95 \, \rm{fm}^{-1}$.
In Fig.~\ref{kcotnlo}, the circles correspond to the phase shift  
analysis of Ref.~\cite{OS67}, 
while the squares show a potential model calculation using the
Argonne V18  nucleon-nucleon potential and  the Urbana three-nucleon force
\cite{KRT96}. The triangle gives the experimental value of the $nd$-scattering
length  from Ref.~\protect\cite{DKN71}. The range corrections are small all the
way up to the breakup threshold. It is encouraging to see that the 
perturbative corrections are small even though $\gamma_t r_t \approx 0.4$ 
is not a very small expansion parameter. This suggests that the EFT 
expansion is well behaved. The range correction clearly improves agreement 
with the phase shift analysis of Ref.~\cite{OS67}. Note that this phase
shift analysis is more than 30 years old and gives no error estimates. 
The errors of the analysis are at least as large as the error of
the scattering length; most likely they are larger.
Consequently, it is more meaningful to compare to the potential
model calculation of \cite{KRT96}, which agrees well with the NLO
result.

In summary, we have calculated the S-wave phase shifts for neutron-deuteron
scattering in the doublet channel to $O(\gamma r)$ 
and found good agreement with available data.  
We have shown that the corrections at this order can be  renormalized by 
modifying the running of the leading order three-body force. Apart
from the two-body effective ranges $r_t$ and $r_s$, no new parameters enter at
this order.  In Refs.~\cite{BeG00,GBG00}, it was shown that the 
perturbative treatment of the range corrections in other channels where 
three-body forces are subleading gives good agreement with available data 
as well. As stated earlier, Ref. \cite{EfT85} has calculated the Phillips 
line to $O(\gamma r)$ and  obtained results which agree well with the 
Phillips line obtained from various potential models. 
Furthermore, if one demands that the neutron-deuteron scattering length
be  correctly reproduced, then the Phillips line predicts the triton binding
energy. The prediction is $8.0$ MeV at $O((\gamma r)^0)$ \cite{BHK00,EfT85}
and $8.8$ MeV at $O(\gamma r)$ \cite{EfT85}. 
These numbers compare well with the measured binding energy of $8.5$ MeV, 
and again the range corrections improve agreement with experiment. 
Together, these results show that the power counting  of
\cite{KSW98,UvK99} is adequate for three-body nuclear systems at very low
energies. This suggests that the perturbative method could be 
used for precise calculations of  phenomenologically important three-body 
processes such as polarization observables in neutron-deuteron scattering 
and the $\beta$-decay  of the triton. 

We acknowledge useful discussions with  P.\ Bedaque,
E.\ Braaten, R.J.\ Furnstahl, R. Perry, and M.\ Strickland. This
work was supported by the National Science Foundation under
Grant No.\ PHY--9800964.

\end{document}